\documentclass[twoside,a4paper,11pt]{sea10}
\usepackage{graphicx}
\usepackage{hyperref}
\usepackage{movie15}
\usepackage{color}
\begin{document}
\pagenumbering{arabic}
\pagestyle{myheadings}
\thispagestyle{empty}
{\flushleft\includegraphics[width=\textwidth,bb=58 650 590 680]{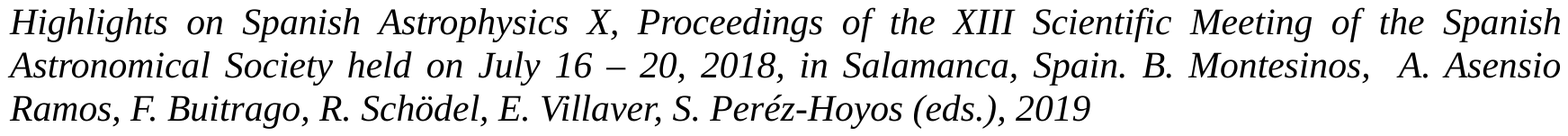}}
\vspace*{0.2cm}
\begin{flushleft}
{\bf {\LARGE
%
\textcolor[rgb]{0,0,1.0}{LiLiMaRlin}, a \textcolor[rgb]{0,0,1.0}{Li}brary of 
\textcolor[rgb]{0,0,1.0}{Li}braries of \textcolor[rgb]{0,0,1.0}{Ma}ssive-Star\linebreak
High-\textcolor[rgb]{0,0,1.0}{R}eso\textcolor[rgb]{0,0,1.0}{l}ut\textcolor[rgb]{0,0,1.0}{i}o\textcolor[rgb]{0,0,1.0}{n} 
Spectra with applications to\linebreak
OWN, MONOS, and CollDIBs
%
}\\
\vspace*{1cm}
%
Jes\'us Ma{\'\i}z Apell\'aniz$^1$, 
Emilio Trigueros P\'aez$^{1,2}$,
Irene Jim\'enez Mart{\'\i}nez$^{1,3}$, 
Rodolfo H. Barb\'a$^4$,
Sergio Sim\'on-D{\'\i}az$^{5}$, 
Anne Pellerin$^6$,
Ignacio Negueruela$^2$, 
and 
Jo\~ao Rodrigo Souza Le\~ao$^7$
%
}\\
\vspace*{0.5cm}
%
$^1$ 
Centro de Astrobiolog{\'\i}a, CSIC-INTA, Spain\\
$^2$ 
Universidad de Alicante, Spain\\
$^3$ 
Universidad Complutense de Madrid, Spain\\
$^4$ 
Universidad de La Serena, Chile\\
$^5$ 
Instituto de Astrof{\'\i}sica de Canarias and Universidad de La Laguna, Spain\\
$^6$ 
State University of New York at Geneseo, U.S.A.\\
$^7$ 
Universidade Federal do Rio Grande do Norte, Brazil\\
%
\end{flushleft}
%
\markboth{
LiLiMaRlin and applications to OWN, MONOS, and CollDIBs
}{ 
%
Ma{\'\i}z Apell\'aniz et al.
%
}
\thispagestyle{empty}
\vspace*{0.4cm}
\begin{minipage}[l]{0.09\textwidth}
\ 
\end{minipage}
\begin{minipage}[r]{0.9\textwidth}
\vspace{1cm}
\section*{Abstract}{\small
%
LiLiMaRlin is a library of libraries of massive-star high-resolution optical spectra built by collecting data 
from [a] our spectroscopic surveys (OWN, IACOB. NoMaDS, and CAF\'E-BEANS) and programs and [b] searches in 
public archives. The current version has 18\,077~spectra of 1665~stars obtained with seven different 
telescopes (HET 9.2~m, NOT 2.56~m, CAHA 2.2~m, MPG/ESO 2.2~m, OHP 1.93~m, Mercator 1.2~m, and Stella 1.2~m).
All the spectra have been filtered to eliminate misidentifications and bad-quality ones, uniformly 
reprocessed, and placed on a common format. We present applications of this library of libraries to the 
analysis of spectroscopic binaries (OWN and MONOS, see poster by E. Trigueros P\'aez at this meeting) and the study 
of the interstellar medium (CollDIBs). We discuss our plans for the future.
%
\normalsize}
\end{minipage}
%
%
%
\section{Introduction}

$\,\!$\indent The last two decades have seen an explosion of stellar high-resolution optical spectroscopic data due to the increase in the 
number of telescopes, some technological improvements, the use of reduction pipelines, and the availability of archives. However, 
significant hurdles remain for the use of the data: telescopes process the data differently and to different degrees, they usually do not 
filter between good- and bad-quality data, and object identification is unclear or has errors in some cases. It is clear that a 
standardization process is needed. Here we present LiLiMaRlin, a \textbf{Li}brary of \textbf{Li}braries of \textbf{Ma}ssive-Star 
High-\textbf{R}eso\textbf{l}ut\textbf{i}o\textbf{n} Spectra which is an ongoing project to achieve that goal for the case of massive stars.

\section{Building LiLiMaRlin}

\begin{table}[ht] 
\caption{Telescopes and data sources used to build LiLiMaRlin.} 
\centerline{
\small
\begin{tabular}{lllccl}
\\
\hline
Telescope       & Spectrograph(s) & Observatory    & Spectral      & Declination      & Source(s)    \\
                &                 &                & resolution(s) & range            &              \\
\hline
HET 9.2 m       & HRS             & McDonald       & 30\,000       & $\ge-$11$^\circ$ & NoMaDS       \\
NOT 2.56 m      & FIES            & La Palma       & 25+46+67\,000 & $\ge-$47$^\circ$ & IACOB+others \\
CAHA 2.2 m      & CAF\'E          & Calar Alto     & 65\,000       & $\ge-$25$^\circ$ & CAF\'E-BEANS \\
MPG-ESO 2.2 m   & FEROS           & La Silla       & 48\,000       & $\le+$25$^\circ$ & OWN+archive  \\
OHP 1.93 m      & ELODIE+SOPHIE   & Haute-Provence & 40+42+75\,000 & $\ge-$24$^\circ$ & archive      \\
Mercator 1.2 m  & HERMES          & La Palma       & 85\,000       & $\ge-$36$^\circ$ & IACOB+others \\
Stella 1.2 m    & SES             & Teide          & 55\,000       & $\ge-$24$^\circ$ & others       \\
\hline
\end{tabular}
}
\label{tab1} 
\end{table}

We build our stellar sample starting from the Galactic O-Star Catalog (GOSC, 
\url{http://gosc.cab.inta-csic.es}, \cite{Maizetal04b,Maizetal12,Maizetal17c,Sotaetal08}), which currently has 8588 objects, most of them
early-type stars. The Galactic O-Star Spectroscopic Survey (GOSSS, \cite{Maizetal11,Maizetal16,Sotaetal11a,Sotaetal14}) is obtaining 
intermediate-resolution spectroscopy of GOSC targets to classify them, of which 2941 have already been observed and processed.

We are building LiLiMaRlin by collecting high-resolution optical spectra obtained with the seven telescopes listed in Table~\ref{tab1}. The
data can be divided into two blocks: (a) Four surveys led by us with the original purpose of studying different aspects of massive stars: 
OWN \cite{Barbetal10}, IACOB \cite{SimDetal15b}, NoMaDS \cite{Maizetal12}, and CAF\'E-BEANS \cite{Neguetal15a}. (b) Spectra found by
searching the public archives and from additional programs led by us for five telescopes: NOT 2.56~m, MPG/ESO 2.2~m, OHP 1.93~m, Mercator 
1.2~m, and Stella 1.2~m.

The data collection is just the first step of the process required to maximize the usefulness of LiLiMaRlin. After having the data, we check
spectrum by spectrum to guarantee the proper source identification, as some archives have data with wrong/unknown IDs or coordinate 
uncertainties of $\sim$2$^\prime$, which are especially problematic in stellar clusters. A related issue is that of source confusion: which 
components of a multiple system are included in the aperture? That requires analyzing target by target with the help of the information in
GOSC, the Washington Double Star catalog (WDS, \cite{Masoetal01}), high-spatial resolution images obtained with the lucky imaging 
instruments AstraLux Norte and Sur \cite{Maiz10a}, and lucky spectroscopy data \cite{Maizetal18a}. We also eliminate noisy, 
lamp-contaminated, or poor order-stitching data.

After the spectra are selected we do our own post-processing. First, we look for errors in the headers by checking the coordinates and time,
the wavelength calibration, and the heliocentric correction. The spectra are then rectified and corrected for telluric lines 
(see \cite{Gardetal13}). Finally, we convert each spectrum into a binary FITS file with a uniform format and generate an entry for the
observation in a MySQL archive that has an interface (based on the one for GOSC: \url{https://gosc.cab.inta-csic.es/gosc-v3-query}) that 
allows us to search the data.

As of the time of this writing, Table~\ref{tab2} lists the data currently in LiLiMaRlin: a total of 18\,077 epochs of 1665~stars.

\begin{table}[ht] 
\caption{Spectra and stars in LiLiMaRlin (October 2017).}
\centerline{
\begin{tabular}{lrrrrr}
\\
\hline
Telescope      & Spectra & Total stars & O stars & \multicolumn{1}{c}{B-A0 stars} & \multicolumn{1}{c}{Other stars}          \\
               &         &             &         & \multicolumn{1}{c}{+ ASG}      & \multicolumn{1}{c}{(WR, late,}           \\
               &         &             &         & \multicolumn{1}{c}{+ sdOB}     & \multicolumn{1}{c}{extragal., unclas.)}  \\
\hline
HET 9.2 m      &     678 &         117 &      74 &                             27 &                                       16 \\
NOT 2.56 m     &    2590 &         639 &     275 &                            203 &                                      161 \\
CAHA 2.2 m     &     834 &         162 &     133 &                             29 &                                        0 \\
MPG-ESO 2.2 m  &    5987 &         675 &     290 &                            302 &                                       83 \\
OHP 1.93 m     &    3952 &         240 &     113 &                            127 &                                        0 \\
Mercator 1.2 m &    3760 &         667 &     166 &                            264 &                                      237 \\
Stella 1.2 m   &     276 &          67 &      56 &                             11 &                                        0 \\
\hline
Total          & 18\,077 &        1665 &     549 &                            691 &                                      425 \\
\hline
\end{tabular}
}
\label{tab2} 
\end{table}



\section{Applications}

$\,\!$\indent We discuss here the two applications we have used LiLiMaRlin for: the calculation of orbits of spectroscopic binaries and the 
analysis of the signature of the intervening ISM in the spectra of hot massive stars.

\subsection{OWN and MONOS: spectroscopic binaries}

$\,\!$\indent The most obvious use of LiLIMaRlin is the calculation of spectroscopic binary (SB) orbits of massive stars, as multiple epochs 
are needed with both good sampling and a large time span (the first epoch in our library dates back to 1994). To study those orbits we have 
divided the sky into a northern and a southern regions, placing the dividing line at $\delta = -20^\circ$ to leave similar numbers of OB 
stars in each region (there are more OB stars in the southern hemisphere due to the position of the Galactic Center, a fact that becomes 
more pronounced at higher extinctions). 

The southern region is the subject of OWN \cite{Barbetal10}, which is based on their own programs with FEROS and
with other spectrographs in Chile and Argentina. There we use LiLiMarlin by supplying additional non-OWN FEROS spectra as well as data
obtained with northern telescopes (we have used the NOT at La Palma to obtain spectra of targets with declinations down to $-47^\circ$).

\begin{figure}
\centerline{\includegraphics[width=0.490\linewidth, bb=28 28 566 566]{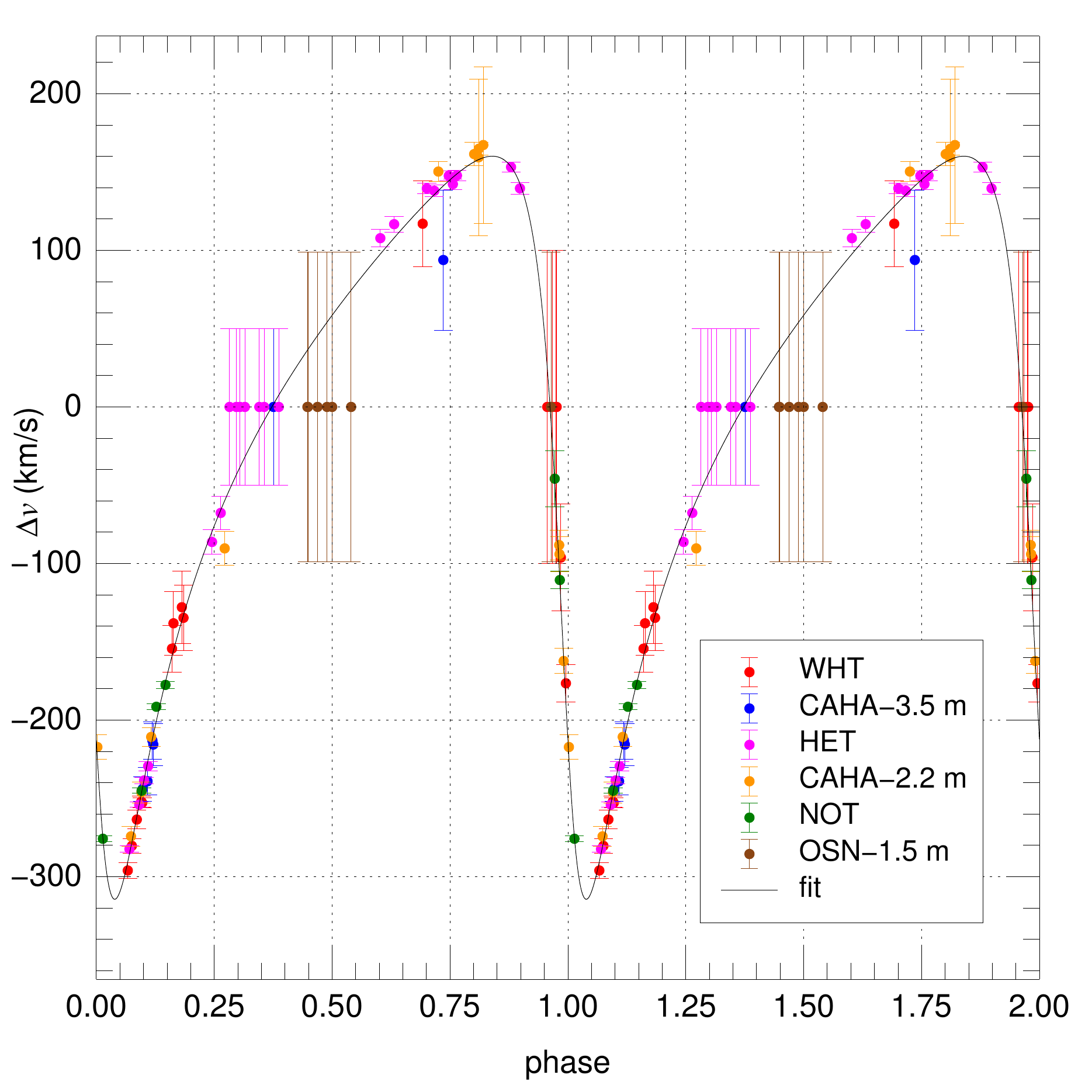} \
            \includegraphics[width=0.490\linewidth, bb=28 28 566 566]{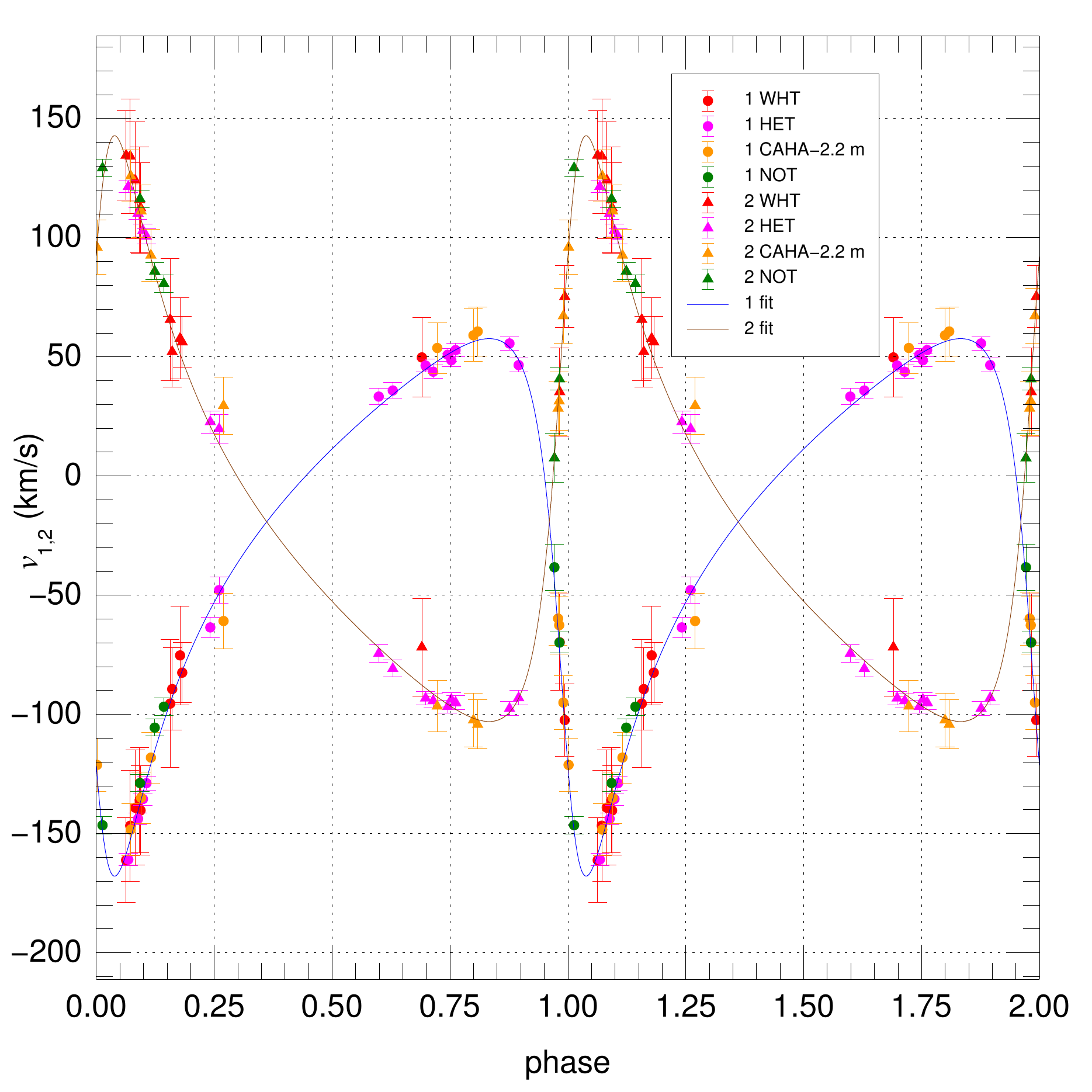}}
\caption{\label{fig1} Phased radial velocity curves for $\Delta v$ (left) and $v_1$ + $v_2$ (right) for the O3.5~If*~+~O3.5~If* SB2 system 
LS~III~+46~11 obtained with LiLiMaRlin and GOSSS data \cite{Maizetal15a}. The color code identifies the telescope.}
\end{figure}

The northern region is the subject of the PhD thesis of Emilio Trigueros P\'aez (see presentation at this meeting) and the 
\textbf{M}ultiplicity \textbf{O}f \textbf{N}orthern \textbf{O}-type \textbf{S}pectroscopic systems (MONOS) project. We have started to 
compare the LiLiMaRlin spectra with the predictions of the publshed orbits in the north and after that we will derive new orbits. The 
prototypes of the MONOS analysis were those of LS~III~$+$46~11 (\cite{Maizetal15a} and Fig.~\ref{fig1}) and $\sigma$~Ori~AaAb
\cite{SimDetal15a}.

The analysis of the spectroscopic orbits is complemented in both the north and the south with information about visual multiples
\cite{Maiz10a,Sotaetal14,Maizetal17a}.

\subsection{CollDIBs: the intervening ISM}

\begin{figure}
\begin{minipage}{0.63\textwidth}
\centerline{\includegraphics[width=\linewidth, bb=28 28 566 566]{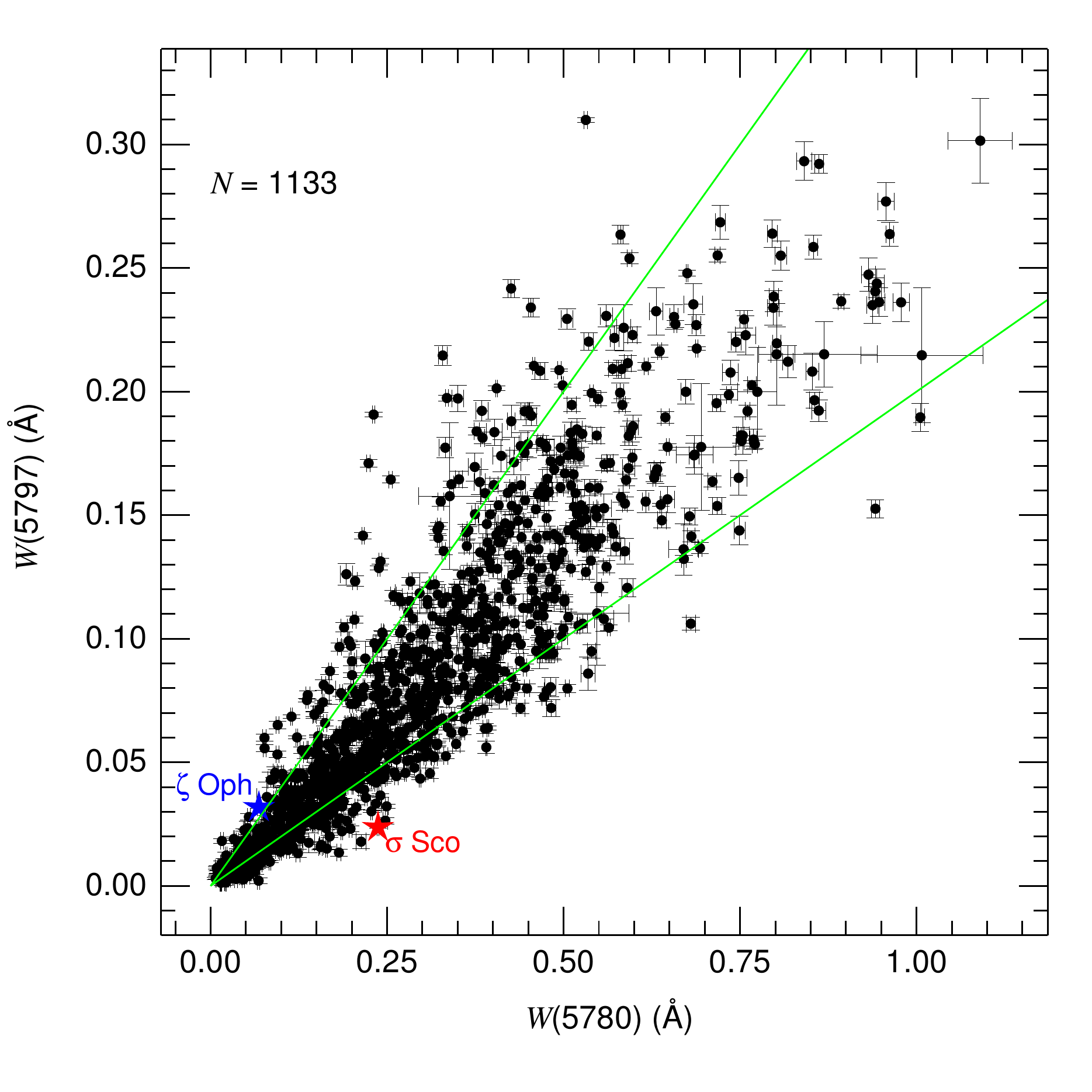}}
\end{minipage}
\begin{minipage}{0.35\textwidth}
\caption{\label{fig2} Equivalent widths of DIB~5780 and DIB~5797 for 1133 CollDIBs sightlines. The two green lines mark the ratios of 
DIB~5797/DIB~5780 of 0.2 and 0.4. The two prototypical sightlines of $\sigma$~Sco and $\zeta$~Oph are indicated.
Compare to the right panel of Fig.~4 in \cite{Maiz15a}, which had only $\sim$1/3 of the data points.}
\end{minipage}
\end{figure}

$\,\!$\indent As a second application, we are using LiLiMaRlin to build \textbf{Coll}ection of \textbf{DIBs} (CollDIBs), the largest 
database of Diffuse Interstellar Bands (DIBs) ever built. The aim of the project is to study several tens of DIBs for over 1000 stars over a 
wide range of extinction values. Some preliminary results are given in \cite{Maizetal15c} and \cite{Maiz15a}. The results will be combined 
with GOSSS spectral types and Gaia distances to map the 3-D distribution of DIBs in the solar neighborhood and the correlations with 
extinction \cite{MaizBarb18}. DIBs are correlated among them but not perfectly: two types of sightlines ($\sigma$ and $\zeta$) are 
characteristic of UV-exposed and UV-shielded ISMs, respectively (Fig.~\ref{fig2}).

\section{Future work}

$\,\!$\indent We will expand LiLiMaRlin by adding spectra from our projects and from the archives we have already explored. We also
plan to explore additional archives such as those of HARPS@ESO~3.6~m, UVES@VLT, ESPaDOnS@CFHT, and NARVAL@Bernard~Lyot. However, our primery
goal is to exploit the LiLiMaRlin data with OWN, MONOS, and CollDIBs.

%
%
\small  
%
\section*{Acknowledgments}   
%
J.M.A. and E.T.P. acknowledge support from the Spanish Government Ministerio de Ciencia, Innovaci\'on y Universidades through grant
AYA2016-75\,931-C2-2-P. E.T.P., S.S.-D., and I.N. acknowledge support from the Spanish Government Ministerio de Ciencia, Innovaci\'on y 
Universidades through grant AYA2015-68\,012-C2-1/2-P. R.H.B. acknowledges support from the ESAC Faculty Council Visitor Program. 
%

%
\end{document}